\documentclass[12pt,preprint]{aastex}
\usepackage{natbib}
\shorttitle{Low ionization state CME plasma}
\shortauthors{Lee and Raymond}

\def\C3{$\rm C~III$ }

\def\Si12{$\rm Si~XII$ }

\begin{document}

\title{Low ionization state plasma in CMEs}

\author{Jin-Yi Lee$^{1}$ and  John C. Raymond$^2$} 
\affil{$^1$Department of Astronomy and Space Science, Kyung Hee University, Yongin, Gyeonggi, 446-701, Republic of Korea \\
$^2$Harvard-Smithsonian Center for Astrophysics, 
Cambridge, MA 02138, USA\\
} 

\begin{abstract}

The Ultraviolet Coronagraph Spectrometer on board the 
{\it {Solar and Heliospheric Observatory (SOHO)}} often observes
low ionization state coronal mass ejection (CME) plasma at ultraviolet wavelengths. 
The CME plasmas are often detected in O~VI (3x10$^5$K), C~III
(8x10$^4$K), Ly$\alpha$, and Ly$\beta$, with the low ionization 
plasma confined to bright filaments or blobs that appear in small 
segments of the UVCS slit.  On the other hand, in situ observations
by the Solar Wind Ion Composition Spectrometer (SWICS) on board 
{\it Advanced Composition Explorer (ACE)} have shown mostly high ionization 
state plasmas in the magnetic clouds in interplanetary coronal mass ejections (ICME)
events, while low ionization states are rarely seen.  In this analysis, we 
investigate whether the low ionization state CME plasmas observed by UVCS 
occupy small enough fractions of the CME to be consistent with the 
small fraction of ACE ICMEs that show low ionization plasma, 
or whether the CME plasma must be further ionized after passing the UVCS slit. 
To do this, we determine the covering factors of low ionization state plasma for 10~CME events. 
We find that the low ionization state plasmas in CMEs observed by UVCS show 
average covering factors below 10\%. 
This indicates that the lack of low ionization state ICME plasmas observed by the ACE 
results from a small probability that the spacecraft passes through a region of low
ionization plasma. 
We also find that the low ionization state plasma covering factors
in faster CMEs are smaller than in slower CMEs. 

\end{abstract}


\clearpage

\section{Introduction}

Coronal mass ejections (CMEs) are among the most explosive 
solar phenomena. It is important to understand the mechanism of 
CME eruption and propagation through interplanetary space, 
which can contribute to understanding the space weather environment.
CMEs are often described as 
a three-part structures: core, cavity, and leading edge
\citep{crifo83, webb88}. The Ultraviolet Coronagraph Spectrometer (UVCS) 
on board the {\it Solar and Heliospheric Observatory (SOHO)} has observed 
CME plasmas in UV spectral lines at a few solar radii in corona. 
The leading edge of CME is often observed in O VI~(1032 \AA) 
and Ly$\alpha$~(1216 \AA) \citep[e.g.][]{raymond03, lee06} as a 
diffuse brightening by a modest factor.  In contrast, the core of 
CME is often observed in relatively low formation temperature lines 
such as C III~(977 \AA) \citep{ciaravella97, ciaravella00, lee09, 
murphy11}, and it is generally confined to a set of very bright 
blobs or filaments that appear in small segments of the UVCS slit. 

\begin{figure}
\epsscale{1.0}\plotone{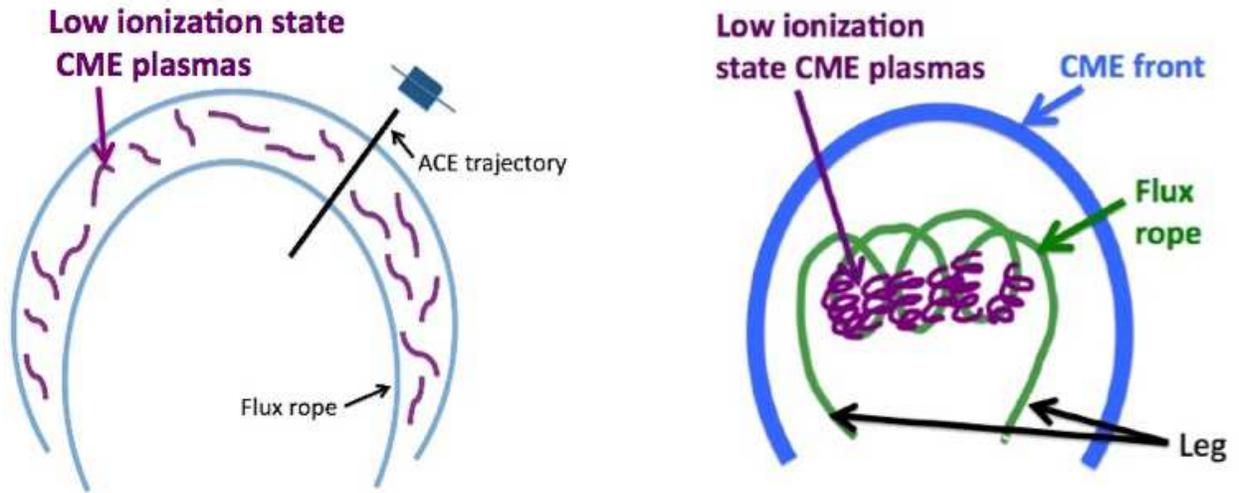}
\caption{Left: Low ionization state plasma in ICME observed by ACE, 
Right: Low ionization state CME plasma observed by UVCS.} 
\label{fig:cme}
\end{figure}

The connection between CMEs and disturbances in the solar wind at 1~AU has been found in 
the comparison of their plasma characteristics such as temperature and ion composition 
\citep[see references in][]{cane2003}.
For example, CMEs can drive interplanetary shocks \citep{sheeley1985}.
Nowadays, the expanded CME structure and its sheath of swept-up solar wind plasma is 
referred to as an Interplanetary Coronal Mass Ejection (ICME) 
\citep[][see also a review for ICME, \citet{howard2009}]{zhao1992, dryer1994}
The Solar Wind Ion Composition Spectrometer (SWICS) on board 
{\it Advanced Composition Explorer (ACE)} observes ICMEs in situ near Lagrangian point~1,
$\sim1.5$  million km from the Earth toward the Sun.  Figure~\ref{fig:cme} 
shows ICME and CME material that might be observed by {\it ACE} and UVCS, respectively.
In CME models, the flux rope may exist before the eruption \citep{lin2000} or 
it may form during the eruption \citep{gosling93}, but in either case it is 
believed to evolve into a smoothly rotating magnetic field structure,
generally considered to be a flux rope or magnetic cloud in interplanetary
space. 

Solar wind ionic charge states become ``frozen-in'' when the ionization and recombination
time scales exceed the expansion time scale, 
i.e. $\tau_{exp} (\equiv \ | \frac {u} {n_e} \frac {\partial n_e} {\partial r} |^{-1}) 
 \ll \tau_{ion}  (\equiv \frac {1} {n_e(C_i+R_i)})
 $ \citep[e.g.][]{hundhausen68, ko99}, where $u$ is velocity, $n_e$ is electron density, 
$C_i$ is ionization rate, and $R_i$ is recombination rate. 
$\tau_{exp}$ and $\tau_{ion}$ are the expansion timescale and  
ionization/recombination timescale, respectively.  
A study using an eclipse observation shows that the freezing in height is
$\sim$1.5$R_\sun$ in the fast solar wind \citep{habbal07}. 
\citet{rakowski07} find that the ionization states of Si and Fe freeze in at 2 to 3 $R_\sun$ 
in models of solar eruptions. The lower ionization state ions we consider here, such as H I, C III, and O VI 
have faster ionization rates and could freeze in at greater heights.

Iron charge state distributions of
ICMEs observed by {\it ACE}/SWICS show that most magnetic cloud plasmas are
highly charged, with ions such as $Fe^{16+}$ \citep{lepri01, lepri04}.  Low ionization
states are relatively rare.
\citet{lepri2010} show that low ionization state plasmas have
been observed only 11 events ($\sim4\%$ of their events) in a
survey of 10 years of SWICS data. 
In addition, they find that
these events originated in filaments near the Sun.
UVCS observations have shown that CME plasma is strongly 
heated even after it leaves the eruption site \citep{akmal01, lee09, murphy11}.
Comparison of time-dependent ionization models with in situ measurements 
at 1 AU also requires strong heating in the region around 
2$R_\sun$ \citep{rakowski07, rakowski11, gruesbeck11, lynch11}. 
Thus it is possible that the low ionization states observed by UVCS are destroyed 
as the plasma moves beyond the height of the UVCS slit.  It is also possible
that much of the low ionization material observed in the corona falls back to
the Sun.  

Thus the question arises why ACE seldom detects low ionization plasma,
while bright, low ionization structures are the salient features of
CMEs in UVCS observations.  One possibility is that it is a matter of
detection criteria.  \citet{lepri2010} used stringent selection criteria
for low ionization plasma.  They used 2 hour binning, and they did not
include singly charged ions.  Considering that the low ionization states
dominate by a large margin in the cool UVCS structures, it is unlikely
that ACE would have seen high ionization states but not low if it passed
through such a structure.  The UVCS structures will expand to several times
$10^{11}$ cm by the time they reach 1 AU, so the time resolution is unlikely
to be an issue.  UVCS generally sees bright C III and O VI, but while C II~(1036 \AA\ and 1037 \AA)
is sometimes observed it has never been found to dominate.  Therefore it
is unlikely that the exclusion of singly charged ions can account for
the difference.  We conclude that the low detection rate of low ionization
plasma in ICMEs compared with CME observations in the corona is not an
artifact of the different measurement techniques.  It must therefore
is result from a small probability that the spacecraft encounters
a clump of low ionization plasma, or else occur because the amount of low ionization plasma
at 1 AU is really smaller than the amount at a few solar radii.

In situ observations by the {\it ACE/SWICS} represent the plasmas
along the {\it ACE} trajectory through the ICME (see the left panel of
Figure~\ref{fig:cme}). 
Assuming that the cool plasma maintains the filamentary structure seen by UVCS, 
the probability that ACE will detect low ionization plasma is proportional to the  
covering factor of the cool filaments, that is the fraction of the 2D projection
of the ICME where cool material is present.
Similarly, the UVCS observations of low ionization state plasma at a few solar radii
can be transformed into 2D images, and the covering factor of low
ionization material equals the probability that a line of sight
passes through a low ionization filament or blob 
(see the right panel of Figure~\ref{fig:cme}). 
In this analysis, we measure the covering factor of low
ionization state plasma observed by {\it SOHO}/UVCS.
This allows us to determine whether the difference
between the low ionization seen by UVCS and the high ionization seen
by {\it ACE}/SWICS results from a small covering factor of cool plasma,
as opposed to heating of CME material as it expands through solar corona or 
draining of cool plasma back to the Sun.

In \S2, we describe the observational data used in this analysis. 
In \S3, we explain how we determine the covering factors of low ionization state CME plasmas observed by UVCS. 
In \S4, we present the covering factors for 10 CME events. 
In \S5, we discuss the results with respect to the heating of CME plasma in lower corona. 
In addition, we examine an independent list of ICME events for which UVCS observes the 
corresponding CME plasma to see what fraction shows low ionization material at coronal heights.

\section{Observations}
\label{sec:obs}

SOHO/UVCS \citep{kohl95} observes the solar corona with an instantaneous 
field of view given by the 40$'$ long
spectrometer entrance slits as projected on the plane of the sky, 
which can be placed between 1.5 $R_\sun$ and 10 $R_\sun$. Different wavelength
ranges are covered in different observations because of tradeoffs with spatial and spectral 
resolution.  We select 10 CME events shown in Table~\ref{tb:prominence}. These 
events have been intensively studied for their kinetic and physical properties 
(see references in Table~\ref{tb:ratio}).
The UVCS slits are placed between 1.4 $R_\sun$ and 2.3 $R_\sun$ for these 
events. The CME speed of the events ranges from about 200 
to 2500 $\rm km~sec^{-1}$, so both slow and fast CME events are included.

First, we select two particularly well observed events on 2000 June 28 and 2000 Oct. 22 
(event numbers 5 and 6 in Table \ref{tb:prominence}).
These two events show the shape of 
erupted prominence material in consecutive UVCS exposures (see \S3.1). 
The observations show very bright emission in the lines 
C III (977 \AA, $8\times10^4$ K) or Ly$\alpha$ (1216 \AA, $< 7\times10^4$ K in ionization equilibrium) 
as well as O VI (1032\AA, $3\times10^5$ K).  The temperatures are formation temperature in 
ionization equilibrium using CHIANTI 7.0 \citep{landi2012}.
The CME plasma may be far from ionization equilibrium due to its rapid expansion speed, 
but the low formation temperatures of these ions indicate that 
the gas was much cooler than coronal temperatures at some point.
For both events, the Extreme Ultraviolet Imaging Telescope 
(EIT) on board SOHO shows a prominence eruption in He II 304 \AA\ on the solar limb.

Second, we select four slow CME events that 
show speeds of 211$-$498 $\rm km~sec^{-1}$ (event numbers 1$-$4 in Table~\ref{tb:prominence}).
All four events are associated with a prominence eruption (see references in Table~\ref{tb:ratio}). 
However, an eruption on 2000 Feb 11 (event number 4) indicates an $H\alpha$ filament 
behind the limb as the most likely source \citep{ciaravella03}. 
A few of the four events are associated with B-class X-ray flares (see Table~\ref{tb:prominence}). 

Lastly, we select four fast CME events that show speeds of 
1913$-$2657 $\rm km~sec^{-1}$ (event numbers 7$-$10 in Table~\ref{tb:prominence}) associated with 
X-class flares. These events occurred in 2002$-$2003 during solar maximum.
EIT 304 \AA\ observations show prominences in the flare occurrence regions associated with these CMEs.
However, the observations were taken every 6 hours, and 
it is not obvious whether the CME events are associated with prominence eruptions or not.
The catalog of prominence and filament in the 
{\it Solar Geophysical Data (SGD)}\footnote[1]{ftp://ftp.ngdc.noaa.gov/STP/SOLAR$\_$DATA/SGD$\_$PDFversion/} 
shows a Loop Prominence System (LPS) in 3 cases (see Table~\ref{tb:prominence}). 
The LPSs are observed later than 
the CME eruption, indicating that the recorded LPS could be a post-flare loop system. 

\section{Analysis}
\label{sec:analysis}

We measure the covering factors of low ionization state CME 
plasmas, defined as the fraction of the reconstructed CME image where
low ionization material is detected, for 10~CME events observed by SOHO/UVCS. First, 
we construct two-dimensional images that show CME material 
along the UVCS slit in consecutive exposures.
Then, we calculate the covering factors of low ionization state plasma 
in an area where CME plasma passes through the UVCS slit in the two-dimensional images.
The UVCS observations did not always cover the full extent of the CME, 
and the events chosen might be biased toward the center of the CME 
where the prominence material is likely to be seen. 
Therefore, there is some tendency for the covering factors obtained from the UVCS observations to be larger than 
would be obtained for the entire CME, but it is probably not a large effect.
 
\subsection{Construction of 2-D image from UVCS observations}

In most UVCS CME-watch observations, the UVCS slit is placed at a fixed position over 
several hours. The observed one-dimensional images of intensity versus position along 
the slit at a single height  
can be placed on a 2-D position-time plane with consecutive exposures. This 2-D image 
represents a temporal 
scanned image of the event at a fixed position. 
The time axis can be multiplied by CME speed to obtain an equivalent spatial image.
These 2-D images can be 
found in the LASCO CME catalogue\footnote[2]{http://cdaw.gsfc.nasa.gov/CME$\_$list/} 
for most of the CME events observed by UVCS.

\begin{figure}
\epsscale{0.8}\plotone{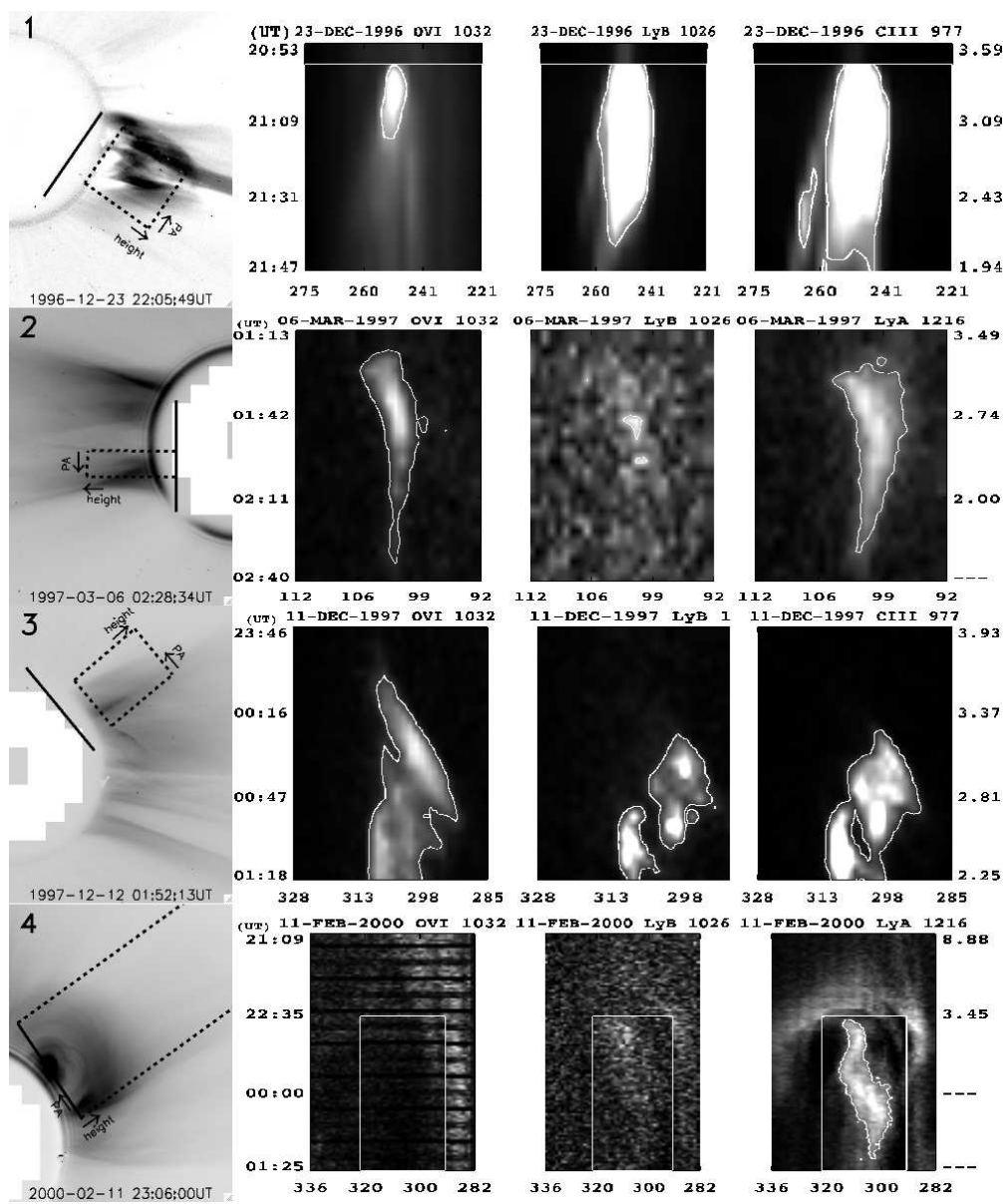}
\caption{LASCO C2 and UVCS observations. The first left panels show the LASCO observations and the locations of the UVCS slit (solid line). 
Black dashed lines (box) represent the location of the UVCS 2$-$D image of the right sides estimated by CME velocities in Table~\ref{tb:prominence}. 
Right three panels show the 2$-$D images of OVI, Ly$\alpha$, Ly$\beta$, or CIII.
Left axis represents the UVCS observation time. Right axis represents the heights estimated by the CME velocities at the UVCS observation time. 
The heights of later UVCS observation than LASCO observation are represented as `$-$$-$$-$'
which indicates that the material with the mark is not yet erupted on the LASCO observation time shown in the left column.
White solid line (box) represents selected area for the covering factor calculation. The events with no white solid line (box) 
encompasses the entire UVCS 2D image.
Contours on 2$-$D images represent selected low ionization CME plasmas. } 
\label{fig:uvcs1}
\end{figure}

\begin{figure}
\epsscale{0.8}\plotone{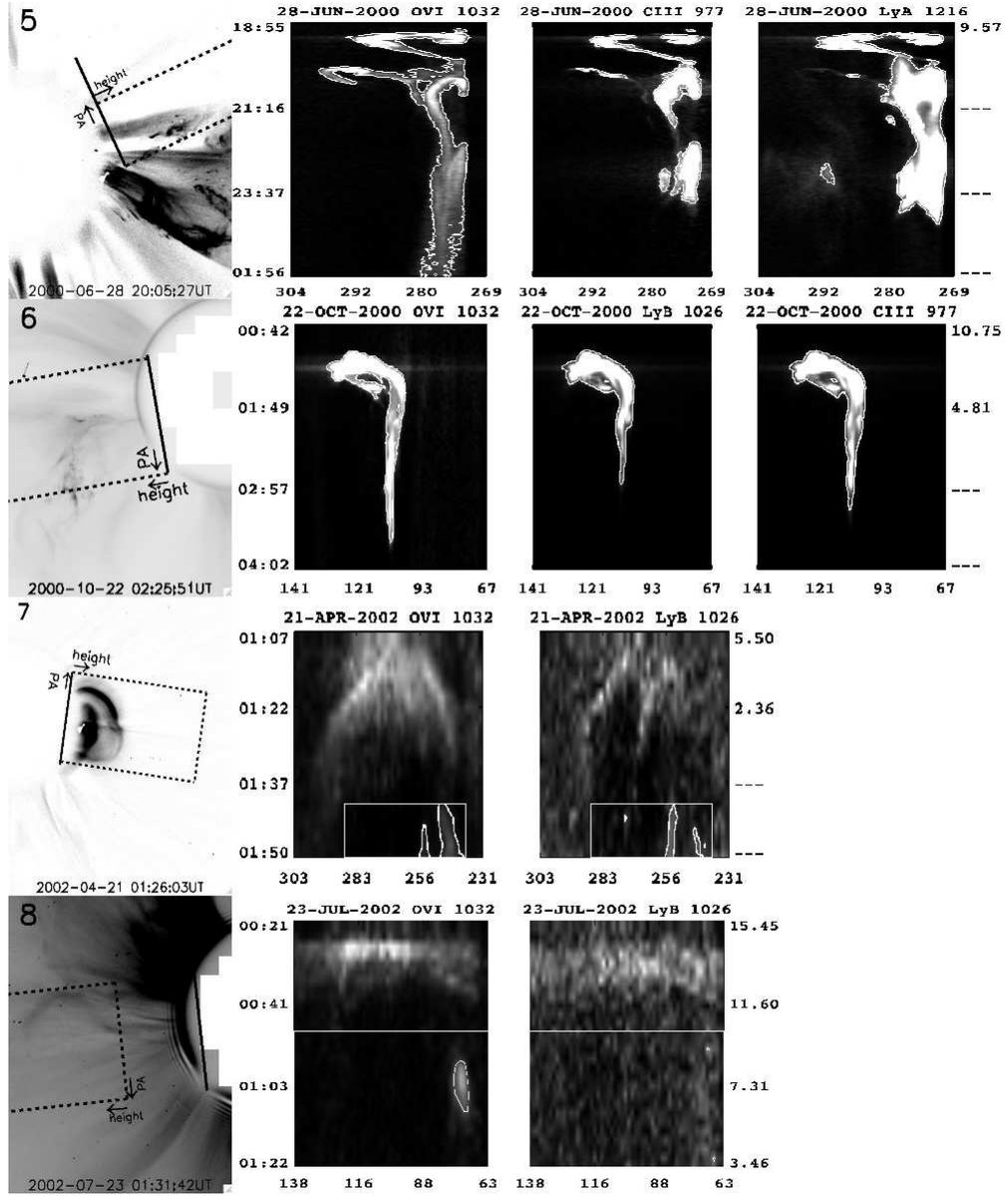}
\caption{Same as Figure~\ref{fig:uvcs1} for four more events. } 
\label{fig:uvcs2}
\end{figure}

\begin{figure}
\epsscale{0.8}\plotone{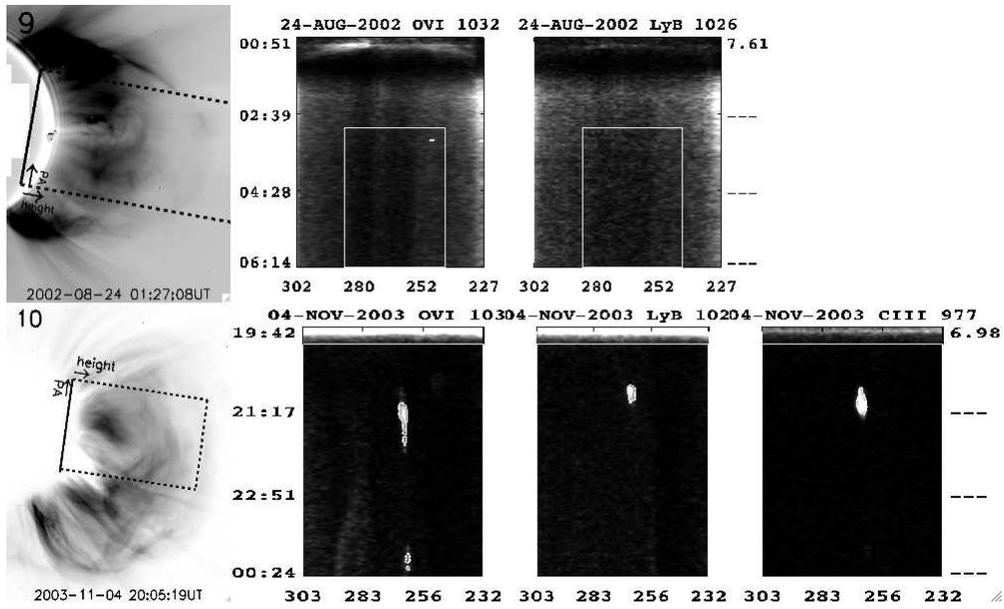}
\caption{Same as Figure~\ref{fig:uvcs1} for the final two events.} 
\label{fig:uvcs3}
\end{figure}

In Figures~\ref{fig:uvcs1} to ~\ref{fig:uvcs3}, we show the 10 CME events used in 
this analysis.  The first column
shows LASCO C2 observations with a solid line segment that represents 
the position of the UVCS slit for each event.
The 2-D images constructed using the observations in O VI, Ly$\alpha$, Ly$\beta$, 
and CIII are placed in the next panels depending on events. 
The dotted box in the first column represents the location of the 2-D images on the LASCO observation.
The horizontal and the left vertical axes represent
polar angles along the slit (counterclockwise from the north pole in degree) and 
observation time, respectively.

For example, an event on 
2000~Oct.~22 helps to understand the constructed 2-D image compared with 
the LASCO observation (see the second row in Figure~\ref{fig:uvcs2}). 
The hook shape in the 2D images can be compared with the eruptive prominence 
on the LASCO observation by rotating of the 2D images  90~$^{\circ}$ counterclockwise. 
  
In addition, we show the heights of CME material corresponding to the height at the time 
observed by UVCS on the 2-D image (the right vertical axis in Figures~\ref{fig:uvcs1} to \ref{fig:uvcs3}). 
The heights are estimated by 
assuming the constant velocity given in CDAW LASCO CME catalog\footnotemark[2]. However, neither the 
speed chosen nor the possibility that the CME is not moving perpendicular to the
UVCS slit affects the measured covering factor.

\subsection{The covering factor of various ionization state CME plasma}

We calculate the covering factors of low ionization 
state CME plasma observed in UV lines, O VI (1032~\AA), Ly$\beta$ (1026 \AA), C III (977~\AA), and 
Ly$\alpha$ (1216~\AA).  UVCS often observes low ionization state plasma 
in the CME core (see a cartoon in the right panel of  Figure~\ref{fig:cme}). 
In this analysis, we exclude the features that indicate CME front and leg because 
these features are probably ambient coronal plasma with the coronal ionization state.
In general the H Lyman lines and O VI lines from the front are diffuse, not very much brighter 
than the pre-CME corona, and their line widths are at least as large as in the 
pre-CME observations. On the other hand, prominence material tends to be 
concentrated in filamentary material, it is very bright, and it often shows a low kinetic temperature 
based on the narrow line width.

To find the covering factor of the low ionization CME plasma, 
we first subtract a background. For 9 events, we took the average of 
1 to 2 hours of pre-CME exposures as the background. 
For the event on Dec.~23~1996, the wavelength setting was changed 
just before the CME occurred, so no background is subtracted. 
Because the prominence emission in low ionization lines is extremely bright 
compared to the pre-CME emission, this has little effect on the results.
A faint emission feature outside of contour later than 21:10UT in O~VI is background emission, so 
the feature was excluded for selecting the low ionization CME plasmas.

Second, we select the area that does not include the CME front and leg.  These are
coronal material that is compressed by the expanding flux rope, and while they
often appear as enhanced emission regions in Ly$\alpha$ and O VI, the ionization
state is that of the ambient corona.
For the 1996~Dec.~23, 2002~Jul.~23 and 2003~Nov.~4 events, we use the area below the white line 
to exclude the background or the CME front.  For the 2000~Feb.~11, 2002~Apr.~21 and 2002~Aug.~24 events, 
we use the area inside the box to exclude the front and leg.  For the other events, we take
the entire reconstructed UVCS image.
We use the same set of spectral lines, O VI, Ly$\beta$, CIII, and Ly$\alpha$ for each event. 
This allows a comparison of the covering factors for material at different formation 
temperatures for each event.

Third, we select the low ionization state plasmas (contours in Figures~\ref{fig:uvcs1} 
to \ref{fig:uvcs3}).
The contour levels are selected with the lowest value which does not include 
background noise features.
Then the covering factors can be calculated by 

\begin{equation}
\mathsf{Covering~factor} = \frac {\mathsf {The~total~number~ of~pixels~in~contoured~area}}  
{\mathsf{The~total~number~of~pixels~in~selected~area}}
\label{eq:ratio}
\end{equation}

The biggest uncertainty in the covering factor comes from the denominator.  
First, we have chosen
the rectangular areas in the UVCS images as large as we can without including emission
from the CME front or legs, but the choice is somewhat subjective. 
It is based upon LASCO movies
rather than the images shown the left hand column in Figures~\ref{fig:uvcs1} to \ref{fig:uvcs3}, 
and the images in the figures can be somewhat misleading.  
In some events, 1996 Dec. 23, 1997 Dec. 12, and 2000 Jun. 28, UVCS observes 
the partial structures of the CMEs because of the slit location. 
We exclude the part of the slit outside of the CME structure.
Second we give the same
weight to each exposure, which is equivalent to assuming a constant speed across the UVCS
slit.  Acceleration would tend make the areas at later times larger, but in general the
expansion speed deep within the CME is smaller than at the front, so the areas at later
times would be diminished.  These effects probably cause an uncertainty at the 30 to 40\%
level, which will not affect our conclusion that the covering factor is small.

\section{Results}
\label{sec:results}

Table~\ref{tb:ratio} shows the covering factors for 10 CME events. A
covering factor of 0.00 means that the low ionization line covering factor was smaller than 0.005.
Events~1$-$4 are slow CMEs with associated prominence eruptions (see 
references for each event in Table~\ref{tb:ratio}). Event~1 shows 
several prominence/filament events (ADF, EPL, DSF) near the CME 
eruption time and location with a small B-class X-ray flare. 
This event was studied as the first SOHO observation of CME 
initiation with a prominence eruption \citep{dere97}.
This event is one of the two events (event~1 and 3) observed in 4 
wavelengths (O VI, Ly$\alpha$, Ly$\beta$, and CIII), which allows the 
comparison of the covering factors for plasma at different formation temperatures.
Event~1 shows higher covering factors in Ly$\beta$, CIII, Ly$\alpha$ than in O VI, 
while in event 3 the O VI covering factor is largest.
Event~2 is likely associated with a B8.9 X-ray flare. 
There is a B-class flare close to the time of event~3. However, the SGD 
shows the flare without its location information, so
it may not be associated with this CME. 
Event 4 is associated with a prominence eruption behind the limb \citep{ciaravella03}.

Events~5 and 6 are associated with prominence eruptions at the solar limb. 
These events are especially well observed events with cool material.
Both events show prominence eruptions in EIT 304 \AA\ observations.
Event~5 was observed to exhibit helical motion as the prominence material 
passed through the UVCS slit \citep{ciaravella05}.  Event~6 especially shows the 
hook shape that provides an easy comparison with the LASCO observation (see \S3.2).
A C-class flare is associated with event~5 while event~6 does not show any associated X-ray flare. 
These CMEs with speeds of $\sim$1000 $\rm km~s^{-1}$ are in the middle of the speed range of the 10 
events.  The covering factor of C III in event~5 is relatively small compared to O VI and Ly$\alpha$.
However, both events show similar covering factors in the different lines. 

Events 7 $-$ 10 are fast CMEs ($\gtrsim$2000 $\rm km~s^{-1}$). All four events are associated 
with X-class flares.  Event~8 shows ejected material in the LASCO observation
that could be the cool material observed in the O VI.  Events 7$-$10 are observed in hot spectral lines \citep{raymond03, ciaravella08}.
In the case of event~10, there are ejecta in EIT 195 \AA\ images \citep{ciaravella08}. 
UVCS observed small blobs of cool material at three times over the course of many hours, suggesting that 
those ejecta arose as result of later magnetic field rearrangement \citep{ciaravella08}. 
We used a time interval that included only the first cool blob in this analysis (Figure~\ref{fig:uvcs3}),
but a similar small covering factor would be obtained with other choices.
All the covering factors are small in all 4 fast CME events.

The slow CME events associated with a prominence/filament show relatively larger fractions of cool 
plasma, while the fast CME events associated with X-class flares show smaller fractions than the 
slow CME events.  This could be because any prominence material in the faster CMEs is more 
strongly heated, so that it is highly ionized before it reaches the height of the UVCS slit.  In
addition, the covering factors at the different formation temperatures for each event are mostly similar. 
Overall, the covering factors in 10 CME events all show small numbers in the range of 0.0$-$0.23.
This indicates that the small number of cool ICME events in ACE observations results 
from a small covering factor of cool plasma.

\section{Discussion and Conclusions}

We show 26 ICME events in Table~\ref{tb:icme}. 
The ICMEs are selected for 1996$-$2002 from \citet{cane2003} and for 2003$-$2005 
from \citet{richardson2010}.  The list also can be found in their ICME 
list\footnote[3]{http://www.ssg.sr.unh.edu/mag/ace/ACElists/ICMEtable.html}.
The list shows a corresponding CME event for each ICME event. 
We select the ICME events in which UVCS observes the corresponding CME plasma from LASCO CME catalogue.
We exclude cases where the corresponding CME is multiple CMEs or a doubtful association. 
Two events show a slightly different CME occurrence time in the ICME list and CME catalogue 
(represented as $^{g,h}$).
The events include 3 cool ICME events in \citet{lepri2010} with a mark $^*$.

In Table~\ref{tb:icme}, about half of events are prominence associated. 
The presence or absence of the associated prominence is indicated 
in the UVCS pages linked to the LASCO CME catalogue.
For a few questionable cases in the catalogue, we examined the UVCS data to determine 
the presence or absence of cool material.
Earlier, it was believed that most CME events are associated with filament/prominence eruption \citep{webb87}.
However, a recent study shows many CMEs are detected without low coronal signatures \citep{ma2010}.
The O VI and Ly$\alpha$ can indicate either the front of the CME and a prominence. 
However, those can be identified by line characteristics (see \S3.2). 
Several events were observed in relatively low temperature lines (e.g. CIII 977 \AA\ ). 
These are represented with a mark $^{**}$.  
One event among these four events is associated with a X-class flare,
while the other three events are associated with C- and small M- class flares.
It is possible that the more energetic flares also have larger heating rates in the
ejected prominence region, so that the prominence gas does not appear in low 
ionization lines in UVCS at coronal heights.

In this analysis, the cool material observed by the UVCS shows a small covering factor, 
indicating that the small number of cool ICME events detected by ACE results from 
a small covering factor of cool plasma. Thus there is no evidence that the prominence material 
must be ionized at heights above the UVCS observations at 1.5$-$2$R_\sun$ in order to 
explain the small fraction of ICMEs that show low ionization material, or that low ionization
plasma drains back to the Sun after passing though the heights observed by UVCS. 
While strong plasma heating is present at these heights \citep{akmal01, lee09, murphy11}, 
the ionization state may be largely frozen-in. 

\acknowledgments

This work was supported by NASA grants NNM07AA02C and NNX09AB17G to the Smithsonian
Astrophysical Observatory, the Korea Meteorological Administration/National Meteorological Satellite Center, and 
the Korea Research Foundation (KRF20100014501).
The CME catalog is generated and maintained at the CDAW Data Center by 
NASA and The Catholic University of America in cooperation with the Naval Research Laboratory. 
SOHO is a project of international cooperation between ESA and NASA. CHIANTI is a collaborative 
project involving the following Universities: Cambridge (UK),
George Mason and Michigan (USA).

\bibliographystyle{apj} 
\bibliography{ms}

\begin{thebibliography}{40}
\expandafter\ifx\csname natexlab\endcsname\relax\def\natexlab#1{#1}\fi

\bibitem[{{Akmal} {et~al.}(2001){Akmal}, {Raymond}, {Vourlidas}, {Thompson},
  {Ciaravella}, {Ko}, {Uzzo}, \& {Wu}}]{akmal01}
{Akmal}, A., {Raymond}, J.~C., {Vourlidas}, A., {Thompson}, B., {Ciaravella},
  A., {Ko}, Y.-K., {Uzzo}, M., \& {Wu}, R. 2001, \apj, 553, 922

\bibitem[{{Cane} \& {Richardson}(2003)}]{cane2003}
{Cane}, H.~V. \& {Richardson}, I.~G. 2003, Journal of Geophysical Research
  (Space Physics), 108, 1156

\bibitem[{{Ciaravella} \& {Raymond}(2008)}]{ciaravella08}
{Ciaravella}, A. \& {Raymond}, J.~C. 2008, \apj, 686, 1372

\bibitem[{{Ciaravella} {et~al.}(1997){Ciaravella}, {Raymond}, {Fineschi},
  {Romoli}, {Benna}, {Gardner}, {Giordano}, {Michels}, {O'Neal}, {Antonucci},
  {Kohl}, \& {Noci}}]{ciaravella97}
{Ciaravella}, A., {Raymond}, J.~C., {Fineschi}, S., {Romoli}, M., {Benna}, C.,
  {Gardner}, L., {Giordano}, S., {Michels}, J., {O'Neal}, R., {Antonucci}, E.,
  {Kohl}, J., \& {Noci}, G. 1997, \apjl, 491, L59+

\bibitem[{{Ciaravella} {et~al.}(2006){Ciaravella}, {Raymond}, \&
  {Kahler}}]{ciaravella06}
{Ciaravella}, A., {Raymond}, J.~C., \& {Kahler}, S.~W. 2006, \apj, 652, 774

\bibitem[{{Ciaravella} {et~al.}(2005){Ciaravella}, {Raymond}, {Kahler},
  {Vourlidas}, \& {Li}}]{ciaravella05}
{Ciaravella}, A., {Raymond}, J.~C., {Kahler}, S.~W., {Vourlidas}, A., \& {Li},
  J. 2005, \apj, 621, 1121

\bibitem[{{Ciaravella} {et~al.}(2001){Ciaravella}, {Raymond}, {Reale},
  {Strachan}, \& {Peres}}]{ciaravella01}
{Ciaravella}, A., {Raymond}, J.~C., {Reale}, F., {Strachan}, L., \& {Peres}, G.
  2001, \apj, 557, 351

\bibitem[{{Ciaravella} {et~al.}(1999){Ciaravella}, {Raymond}, {Strachan},
  {Thompson}, {Cyr}, {Gardner}, {Modigliani}, {Antonucci}, {Kohl}, \&
  {Noci}}]{ciaravella99}
{Ciaravella}, A., {Raymond}, J.~C., {Strachan}, L., {Thompson}, B.~J., {Cyr},
  O.~C.~S., {Gardner}, L., {Modigliani}, A., {Antonucci}, E., {Kohl}, J., \&
  {Noci}, G. 1999, \apj, 510, 1053

\bibitem[{{Ciaravella} {et~al.}(2000){Ciaravella}, {Raymond}, {Thompson}, {van
  Ballegooijen}, {Strachan}, {Li}, {Gardner}, {O'Neal}, {Antonucci}, {Kohl}, \&
  {Noci}}]{ciaravella00}
{Ciaravella}, A., {Raymond}, J.~C., {Thompson}, B.~J., {van Ballegooijen}, A.,
  {Strachan}, L., {Li}, J., {Gardner}, L., {O'Neal}, R., {Antonucci}, E.,
  {Kohl}, J., \& {Noci}, G. 2000, \apj, 529, 575

\bibitem[{{Ciaravella} {et~al.}(2003){Ciaravella}, {Raymond}, {van
  Ballegooijen}, {Strachan}, {Vourlidas}, {Li}, {Chen}, \&
  {Panasyuk}}]{ciaravella03}
{Ciaravella}, A., {Raymond}, J.~C., {van Ballegooijen}, A., {Strachan}, L.,
  {Vourlidas}, A., {Li}, J., {Chen}, J., \& {Panasyuk}, A. 2003, \apj, 597,
  1118

\bibitem[{{Crifo} {et~al.}(1983){Crifo}, {Picat}, \& {Cailloux}}]{crifo83}
{Crifo}, F., {Picat}, J.~P., \& {Cailloux}, M. 1983, \solphys, 83, 143

\bibitem[{{Dere} {et~al.}(1997){Dere}, {Brueckner}, {Howard}, {Koomen},
  {Korendyke}, {Kreplin}, {Michels}, {Moses}, {Moulton}, {Socker}, {St.~Cyr},
  {Delaboudini{\`e}re}, {Artzner}, {Brunaud}, {Gabriel}, {Hochedez}, {Millier},
  {Song}, {Chauvineau}, {Marioge}, {Defise}, {Jamar}, {Rochus}, {Catura},
  {Lemen}, {Gurman}, {Neupert}, {Clette}, {Cugnon}, {van Dessel}, {Lamy},
  {Llebaria}, {Schwenn}, \& {Simnett}}]{dere97}
{Dere}, K.~P., {Brueckner}, G.~E., {Howard}, R.~A., {Koomen}, M.~J.,
  {Korendyke}, C.~M., {Kreplin}, R.~W., {Michels}, D.~J., {Moses}, J.~D.,
  {Moulton}, N.~E., {Socker}, D.~G., {St.~Cyr}, O.~C., {Delaboudini{\`e}re},
  J.~P., {Artzner}, G.~E., {Brunaud}, J., {Gabriel}, A.~H., {Hochedez}, J.~F.,
  {Millier}, F., {Song}, X.~Y., {Chauvineau}, J.~P., {Marioge}, J.~P.,
  {Defise}, J.~M., {Jamar}, C., {Rochus}, P., {Catura}, R.~C., {Lemen}, J.~R.,
  {Gurman}, J.~B., {Neupert}, W., {Clette}, F., {Cugnon}, P., {van Dessel},
  E.~L., {Lamy}, P.~L., {Llebaria}, A., {Schwenn}, R., \& {Simnett}, G.~M.
  1997, \solphys, 175, 601

\bibitem[{{Dryer}(1994)}]{dryer1994}
{Dryer}, M. 1994, \ssr, 67, 363

\bibitem[{{Gosling}(1993)}]{gosling93}
{Gosling}, J.~T. 1993, \jgr, 98, 18937

\bibitem[{{Gruesbeck} {et~al.}(2011){Gruesbeck}, {Lepri}, {Zurbuchen}, \&
  {Antiochos}}]{gruesbeck11}
{Gruesbeck}, J.~R., {Lepri}, S.~T., {Zurbuchen}, T.~H., \& {Antiochos}, S.~K.
  2011, \apj, 730, 103

\bibitem[{{Habbal} {et~al.}(2007){Habbal}, {Morgan}, {Johnson}, {Arndt}, {Daw},
  {Jaeggli}, {Kuhn}, \& {Mickey}}]{habbal07}
{Habbal}, S.~R., {Morgan}, H., {Johnson}, J., {Arndt}, M.~B., {Daw}, A.,
  {Jaeggli}, S., {Kuhn}, J., \& {Mickey}, D. 2007, \apj, 663, 598

\bibitem[{{Howard} \& {Tappin}(2009)}]{howard2009}
{Howard}, T.~A. \& {Tappin}, S.~J. 2009, \ssr, 147, 31

\bibitem[{{Hundhausen} {et~al.}(1968){Hundhausen}, {Gilbert}, \&
  {Bame}}]{hundhausen68}
{Hundhausen}, A.~J., {Gilbert}, H.~E., \& {Bame}, S.~J. 1968, \apjl, 152, L3+

\bibitem[{{Ko} {et~al.}(1999){Ko}, {Gloeckler}, {Cohen}, \& {Galvin}}]{ko99}
{Ko}, Y., {Gloeckler}, G., {Cohen}, C.~M.~S., \& {Galvin}, A.~B. 1999, \jgr,
  104, 17005

\bibitem[{{Kohl} {et~al.}(1995){Kohl}, {Esser}, {Gardner}, {Habbal},
  {Daigneau}, {Dennis}, {Nystrom}, {Panasyuk}, {Raymond}, {Smith}, {Strachan},
  {van Ballegooijen}, {Noci}, {Fineschi}, {Romoli}, {Ciaravella}, {Modigliani},
  {Huber}, {Antonucci}, {Benna}, {Giordano}, {Tondello}, {Nicolosi}, {Naletto},
  {Pernechele}, {Spadaro}, {Poletto}, {Livi}, {von der L{\"u}he}, {Geiss},
  {Timothy}, {Gloeckler}, {Allegra}, {Basile}, {Brusa}, {Wood}, {Siegmund},
  {Fowler}, {Fisher}, \& {Jhabvala}}]{kohl95}
{Kohl}, J.~L., {Esser}, R., {Gardner}, L.~D., {Habbal}, S., {Daigneau}, P.~S.,
  {Dennis}, E.~F., {Nystrom}, G.~U., {Panasyuk}, A., {Raymond}, J.~C., {Smith},
  P.~L., {Strachan}, L., {van Ballegooijen}, A.~A., {Noci}, G., {Fineschi}, S.,
  {Romoli}, M., {Ciaravella}, A., {Modigliani}, A., {Huber}, M.~C.~E.,
  {Antonucci}, E., {Benna}, C., {Giordano}, S., {Tondello}, G., {Nicolosi}, P.,
  {Naletto}, G., {Pernechele}, C., {Spadaro}, D., {Poletto}, G., {Livi}, S.,
  {von der L{\"u}he}, O., {Geiss}, J., {Timothy}, J.~G., {Gloeckler}, G.,
  {Allegra}, A., {Basile}, G., {Brusa}, R., {Wood}, B., {Siegmund}, O.~H.~W.,
  {Fowler}, W., {Fisher}, R., \& {Jhabvala}, M. 1995, \solphys, 162, 313

\bibitem[{{Landi} {et~al.}(2012){Landi}, {Del Zanna}, {Young}, {Dere}, \&
  {Mason}}]{landi2012}
{Landi}, E., {Del Zanna}, G., {Young}, P.~R., {Dere}, K.~P., \& {Mason}, H.~E.
  2012, \apj, 744, 99

\bibitem[{{Lee} {et~al.}(2006){Lee}, {Raymond}, {Ko}, \& {Kim}}]{lee06}
{Lee}, J., {Raymond}, J.~C., {Ko}, Y., \& {Kim}, K. 2006, \apj, 651, 566

\bibitem[{{Lee} {et~al.}(2009){Lee}, {Raymond}, {Ko}, \& {Kim}}]{lee09}
---. 2009, \apj, 692, 1271

\bibitem[{{Lepri} \& {Zurbuchen}(2004)}]{lepri04}
{Lepri}, S.~T. \& {Zurbuchen}, T.~H. 2004, Journal of Geophysical Research
  (Space Physics), 109, 1112

\bibitem[{{Lepri} \& {Zurbuchen}(2010)}]{lepri2010}
---. 2010, \apjl, 723, L22

\bibitem[{{Lepri} {et~al.}(2001){Lepri}, {Zurbuchen}, {Fisk}, {Richardson},
  {Cane}, \& {Gloeckler}}]{lepri01}
{Lepri}, S.~T., {Zurbuchen}, T.~H., {Fisk}, L.~A., {Richardson}, I.~G., {Cane},
  H.~V., \& {Gloeckler}, G. 2001, \jgr, 106, 29231

\bibitem[{{Lin} \& {Forbes}(2000)}]{lin2000}
{Lin}, J. \& {Forbes}, T.~G. 2000, \jgr, 105, 2375

\bibitem[{{Lynch} {et~al.}(2011){Lynch}, {Reinard}, {Mulligan}, {Reeves},
  {Rakowski}, {Allred}, {Li}, {Laming}, {MacNeice}, \& {Linker}}]{lynch11}
{Lynch}, B.~J., {Reinard}, A.~A., {Mulligan}, T., {Reeves}, K.~K., {Rakowski},
  C.~E., {Allred}, J.~C., {Li}, Y., {Laming}, J.~M., {MacNeice}, P.~J., \&
  {Linker}, J.~A. 2011, \apj, 740, 112

\bibitem[{{Ma} {et~al.}(2010){Ma}, {Attrill}, {Golub}, \& {Lin}}]{ma2010}
{Ma}, S., {Attrill}, G.~D.~R., {Golub}, L., \& {Lin}, J. 2010, \apj, 722, 289

\bibitem[{{Mancuso} \& {Avetta}(2008)}]{mancuso08}
{Mancuso}, S. \& {Avetta}, D. 2008, \apj, 677, 683

\bibitem[{{Murphy} {et~al.}(2011){Murphy}, {Raymond}, \& {Korreck}}]{murphy11}
{Murphy}, N.~A., {Raymond}, J.~C., \& {Korreck}, K.~E. 2011, ArXiv e-prints

\bibitem[{{Rakowski} {et~al.}(2007){Rakowski}, {Laming}, \&
  {Lepri}}]{rakowski07}
{Rakowski}, C.~E., {Laming}, J.~M., \& {Lepri}, S.~T. 2007, \apj, 667, 602

\bibitem[{{Rakowski} {et~al.}(2011){Rakowski}, {Laming}, \&
  {Lyutikov}}]{rakowski11}
{Rakowski}, C.~E., {Laming}, J.~M., \& {Lyutikov}, M. 2011, \apj, 730, 30

\bibitem[{{Raymond} {et~al.}(2003){Raymond}, {Ciaravella}, {Dobrzycka},
  {Strachan}, {Ko}, {Uzzo}, \& {Raouafi}}]{raymond03}
{Raymond}, J.~C., {Ciaravella}, A., {Dobrzycka}, D., {Strachan}, L., {Ko}, Y.,
  {Uzzo}, M., \& {Raouafi}, N. 2003, \apj, 597, 1106

\bibitem[{{Raymond} {et~al.}(2007){Raymond}, {Holman}, {Ciaravella},
  {Panasyuk}, {Ko}, \& {Kohl}}]{raymond07}
{Raymond}, J.~C., {Holman}, G., {Ciaravella}, A., {Panasyuk}, A., {Ko}, Y., \&
  {Kohl}, J. 2007, \apj, 659, 750

\bibitem[{{Richardson} \& {Cane}(2010)}]{richardson2010}
{Richardson}, I.~G. \& {Cane}, H.~V. 2010, \solphys, 264, 189

\bibitem[{{Sheeley} {et~al.}(1985){Sheeley}, {Howard}, {Michels}, {Koomen},
  {Schwenn}, {Muehlhaeuser}, \& {Rosenbauer}}]{sheeley1985}
{Sheeley}, Jr., N.~R., {Howard}, R.~A., {Michels}, D.~J., {Koomen}, M.~J.,
  {Schwenn}, R., {Muehlhaeuser}, K.~H., \& {Rosenbauer}, H. 1985, \jgr, 90, 163

\bibitem[{{Webb}(1988)}]{webb88}
{Webb}, D.~F. 1988, \jgr, 93, 1749

\bibitem[{{Webb} \& {Hundhausen}(1987)}]{webb87}
{Webb}, D.~F. \& {Hundhausen}, A.~J. 1987, \solphys, 108, 383

\bibitem[{{Zhao}(1992)}]{zhao1992}
{Zhao}, X. 1992, \jgr, 97, 15051

\end{thebibliography}

\clearpage

\begin{deluxetable}{cllllllll} \tabletypesize{\scriptsize}
\tablecaption{CME and Prominence/Filament} 
\tablewidth{0pt}
\tablehead{ \colhead{event} & \colhead{Date} & \colhead{Time$^a$} &
  \colhead{Speed$^a$} & \colhead{PA$^a$} & \colhead{Width$^a$} &
  \colhead{NOAA} & \colhead{GOES X-ray flare$^b$} & \colhead{Prominence/Filament$^b$}  }

\startdata 
1 & 1996 Dec 23 & 21:16 &  354 &  255 & 58 & 8005 & B2$^c$ & ADF$~~$20:06$-$20:11 \\ 
 &  & & &  &   &   &  & EPL$~~$20:11$-$20:48   \\ 
 &  & & &  &   &   &  & DSF$~~$20:16$-$20:46   \\ 
2 & 1997 Mar  6 & 01:36 &  301 & 104 & 27 & 8020 &
B8.9$~~$00:41$-$00:52$~~$N02E78$^d$ & ASR$~~$00:45$-$10:30  \\
3 &1997 Dec 12 & 01:27 &  211 & 291 & 80 &    & B5.2$~~$00:44$-$01:00$~~$ & None  \\
4 & 2000 Feb 11 & 21:08 &  498 & 277 & $>$173 &      & None &
None  \\
5 & 2000 Jun 28 & 19:31 & 1198 &  270 & $>$134 & 9046 & C3.7$~~$18:48$-$19:10$~~$N20W9 &
EPL$~~$18:31$-$20:49  \\
6 & 2000 Oct 22 & 00:50 & 1024 &  103 & 236 &   & None &
EPL$~~$22:30(10/21)$-$01:17  \\
7 & 2002 Apr 21 & 01:27 & 2393 &  Halo & 360 & 9906 & X1.5$~~$00:43$-$02:38$~~$S14W84  &
LPS$~~$02:01$-$09:56 \\
8 & 2002 Jul 23 & 00:42 & 2285 &  Halo & 360 & 10039 & X4.8$~~$00:18$-$00:47$~~$S13E72 &
None \\
9 & 2002 Aug 24 & 01:27 & 1913 &  Halo & 360 & 10069 & X3.1$~~$00:49$-$01:12$~~$S02W81 &
LPS$~~$01:12$-$07:20 \\
10 & 2003 Nov  4 & 19:54 & 2657 &  Halo & 360 & 10486 & X28 $~~$19:29$-$20:06$~~$S19W83
& LPS$~~$21:06$-$00:00 \\
\enddata 
\tablecomments{
{$^a$}: Time (UT), Linear speed (km/s), PA (deg), and Angular width (deg) in the CME catalog (http://cdaw.gsfc.nasa.gov/CME$\_$list) \\
{$^b$}: Solar Geophysical Data(http://www.ngdc.noaa.gov/stp/solar/sgd.html). 
ADF:~Active Dark Filament, EPL:~Eruptive Prominence on the Limb, DSF:~Disappearing Solar Filament,
ASR:~Active Surge Region, LPS:~Loop Prominence System \\
{$^c$}: see \citet{dere97}, No X-ray flare in the Solar Geophysical Data \\
{$^d$}: NOAA active region location
 }
\label{tb:prominence}
\end{deluxetable}

\begin{deluxetable}{clllllllll} \tabletypesize{\scriptsize}
\tablecaption{Covering factor of low ionization CME plasma} 
\tablewidth{0pt}
\tablehead{ \colhead{event} & \colhead{Date} &  
 \colhead{PA(deg)$^a$} &
\colhead{h(R$_\sun$)$^a$} &
  \colhead{OVI} & \colhead{Ly$\beta$} &
  \colhead{CIII} & \colhead{Ly$\alpha$} &
  \colhead{Ref.$^b$} }

\startdata 
1 & 1996 Dec 23 & 235 & 1.39 & 0.03 & 0.18 & 0.23 & 0.36 & 1 \\ 
2 & 1997 Mar  6 &  90 & 1.55 & 0.11 & 0.01 &  *   & 0.13 &  2 \\
3 &1997 Dec 12 & 310 & 1.63 & 0.20 & 0.10 & 0.16 & 0.14 & 3, 4 \\
4 & 2000 Feb 11 &  305 & 2.33 & 0.00 & 0.00 &  *   & 0.24 &  5 \\
5 & 2000 Jun 28 &  295 & 2.32 & 0.15 &  *   & 0.08 & 0.16 &  6, 7 \\
6 & 2000 Oct 22 & 100 & 1.63 & 0.06 & 0.06 & 0.07 &  *   &  7 \\
7 & 2002 Apr 21 & 262 & 1.63 & 0.14 & 0.07 &  *   &  *   & 7, 8, 9\\
8 & 2002 Jul 23 & 96 & 1.63 & 0.02 & 0.00 &  *   &  *   & 8, 10, 11\\
9 & 2002 Aug 24 & 260 & 1.63 & 0.00 & 0.00 &  *   &  *   & 8 \\
10 & 2003 Nov  4 & 262 & 1.63 & 0.01 & 0.00 & 0.00 &  *   & 12 \\

\enddata 
\tablecomments{{*}: No UVCS observation in the wavelength ranges \\
{$^a$}: UVCS slit position angle (PA) and height (h) \\
$^b$ References: 1; \citet{ciaravella97}, 2; \citet{ciaravella99}, 3;
\citet{ciaravella00}, 4; \citet{ciaravella01}, 5;
\citet{ciaravella03}, 6; \citet{ciaravella05}, 7;
\citet{ciaravella06}, 8; \citet{raymond03}, 9; \citet{lee06},
10; \citet{raymond07}, 11;\citet{mancuso08},
12; \citet{ciaravella08} }
\label{tb:ratio}
\end{deluxetable}

\begin{deluxetable}{lllllllll} \tabletypesize{\scriptsize}
\tablecaption{ICMEs \citep{cane2003, richardson2010} with UVCS observations} 
\tablewidth{0pt}
\tablehead{ \colhead{Disturbance$^a$} & \colhead{CME$^a$} &  
 \colhead{Vel.$^b$} & \colhead{Flare$^c$} &
\colhead{PA} & \colhead{h(R$_\sun$)$^d$} &
  \colhead{UVCS obs.} & \colhead{P$^e$} & \colhead{Note$^f$}}

\startdata 
1997 02/09 1321 & 02/07 0030 & 490 & None & 270 & 1.5$-$3.0 & OVI & N & L$?$ \\
$^*$1998 05/01 2156 & 04/29 1658 & 1374 & M6.8 & 144 & 1.9$-$3.8 & Ly$\alpha$ & Y &V$?$\\
1998 11/07 0815 & 11/04 0418$^g$ & 102 &  C5.2 & 359 &  3.1, 3.6 & Ly$\alpha$ & N & \\
1999 07/06 1509 & 07/03 1954 & 536 & C5.6 & 360 & 6.1 & Ly$\alpha$ & N & L \\
2000 01/22 0023 & 01/18 1754 & 739 & M3.9 & 255 & 1.6, 1.9 & Ly$\alpha$, OVI & N & F \\
2000 02/11 0258 & 02/08 0930 & 1079 & M1.3 & 102 & 2.3, 2.6 & Ly$\alpha$ & N & F \\
2000 02/11 2352 & 02/10 0230 & 944 & C7.3 & 102, 110 &1.9,  2.3 & OVI & N & V$?$, L$?$ \\
2000 02/14 0731 & 02/12 0431 & 1107 & M1 & 305 & 2.3 & Ly$\alpha$ & N & F, V$?$ \\
$^{**}$2000 04/06 1639 & 04/04 1632 & 1188 & C9 & 225 & 1.4, 1.5 & Ly$\alpha$, Ly$\beta$, CIII, OVI, NIII & Y & L \\
$^*$2000 07/15 1437 & 07/14 1054 & 1674 & X5.7 & 180 & 1.6$-$4.0 & Ly$\alpha$ & Y & V$?$ \\
2001 03/03 1121 & 02/28 1450 &  313 & None & 225 & 3.1, 2.6 & OVI & Y & F$?$, L$?$ \\
2001 03/27 1747 & 03/25 1706 & 677 & C9 & 360 & 3.1 & Ly$\alpha$ & N & F, S$?$ \\
2001 04/04 1455 & 04/02 2206 & 2505 & X20 & 223, 225 & 2.0, 2.5 & Ly$\alpha$, OVI & Y & F, S$?$ \\
2001 04/11 1343 & 04/10 0530 & 2411 & X2.3 & 270 & 2.6 & Ly$\alpha$ & N & F \\
2001 08/17 1103 & 08/14 1601 & 618 & None & 26 & 2.0 & OVI & Y & F, S$?$ \\
$^{**}$2001 10/11 1701 & 10/09 1130 & 973 & M1.4 & 90$-$180 & 1.9$-$3.1 & Ly$\alpha$, Ly$\beta$, CIII, OVI & Y & F \\
$^{**}$2001 11/19 1815 & 11/17 0530 & 1379 & M2.8 & 0$-$135 & 1.7, 1.5 & Ly$\alpha$, Ly$\beta$, OVI, SiIII, NIII & Y & \\
2001 11/24 0656 & 11/22 2330 & 1437 & M9.9 & 356 & 2.4 & Ly$\alpha$, OVI & Y & F$?$\\
2002 05/23 1050 & 05/22 0326$^h$ &  1557 & C5.0 & 180, 225 & 1.5$-$1.7 & Ly$\alpha$, OVI & Y & \\
$^{**}$2002 07/17 1603 & 07/15 2030 & 1151 & X3.0 & 360 & 1.7$-$3.6 & Ly$\alpha$, Ly$\beta$, CIII, OVI  & Y &  F, S, V$?$ \\
2003 05/29 1825 & 05/28 0050 & 1366 & X3.6 & 360 & 1.6, 1.7 & Ly$\alpha$, OVI & N & \\
$^*$2003 10/28 0206 & 10/26 1754 & 1537 & X1.2 & 245$-$270 & 1.7$-$3.1 & Ly$\alpha$, OVI & N & F, S$?$ \\
2003 10/29 0611 & 10/28 1130 & 2459 & X17.2 & 90$-$225 & 1.7$-$3.1 & Ly$\alpha$, OVI &  Y & F, S$?$\\
2003 10/30 1619 & 10/29 2054 & 2029 & X11 & 178, 179 & 2.0, 2.5 & Ly$\alpha$, OVI & N & F$?$, L$?$ \\
2005 01/21 1714 & 01/20 0654 & 882 & X7.1 & 283 & 2.3  & Ly$\alpha$ & N & \\
2005 05/29 0905 & 05/26 1506 & 586 & B7.5$?$ & 270 & 3.0, 2.1 & FeXVIII, OVI & N & \\

\enddata 
\tablecomments{
$^*$: Cold ICME events in \citet{lepri2010} \\
$^{**}$: Relatively low temperature line observations by UVCS.\\
$^a$: The time of associated geomagnetic storm sudden commencement or shock passage in the ICME events and the associated LASCO CME events \citep{cane2003, richardson2010}.  \\
$^b$: CME speed in the LASCO CME catalog (linear speed km/s) \\
$^c$: GOES X-ray flare. It is referred from log files in the CME catalogue and X-ray flare data in SGD. The last one represented with $?$ shows no flare location information in the SGD. \\
$^d$: CME plasma detected height. If it is not specified in the catalog, it is the UVCS slit height.\\
$^e$: Prominence \\
$^f$: L: Leg, F: Front, S: Shock, V: Void \\
$^g$: LASCO CME 11/04 0454\\
$^h$: LASCO CME 05/22 0350
}

\label{tb:icme}
\end{deluxetable}

\end{document}